\documentclass[aps,prd,groupedaddress,showpacs]{revtex4}

\usepackage[parfill]{parskip}    % Activate to begin paragraphs with an empty line rather than an indent
\usepackage{graphicx}
\usepackage{amssymb}
\usepackage{epstopdf}
\usepackage{color}
\usepackage{float}
\usepackage{amsmath}
\usepackage{mathrsfs}  

\usepackage{graphicx}
\usepackage{pdfpages}
\usepackage[colorlinks=true,citecolor=blue,urlcolor=magenta,breaklinks]{hyperref}
\begin{document}

\title{Wave zone of the Ho\v{r}ava--Lifshitz gravity coupled to a gauge vector }

\author{J. Mestra-Páez}
\email{jarvin.mestra@ua.cl}
\affiliation{Departamento de F\'isica, Facultad de Ciencias Básicas, Universidad de Antofagasta, Casilla 170, Antofagasta, Chile.}

\author{Alvaro Restuccia}
\email{alvaro.restuccia@uantof.cl}
\affiliation{Departamento de F\'isica, Facultad de Ciencias Básicas, Universidad de Antofagasta, Casilla 170, Antofagasta, Chile.}

\author{Francisco Tello-Ortiz}
\email{francisco.tello@ua.cl}
\affiliation{Departamento de F\'isica, Facultad de Ciencias Básicas, Universidad de Antofagasta, Casilla 170, Antofagasta, Chile.}

\begin{abstract}
We consider the anisotropic gravity--gauge vector coupling in the non--projectable Ho\v{r}ava-Lifshitz theory at the kinetic conformal point, in the low energy regime. We show that the canonical formulation of the theory, evaluated at its constraints, reduces to a canonical formulation solely in terms of the physical degrees of freedom. The corresponding reduced Hamilton defines the ADM energy of the system. We obtain its explicit expression and discuss its relation to the ADM energy of the Einstein-Maxwell theory. We then show that there exists, in this theory, a well--defined wave zone. In it, the physical degrees of freedom ı.e., the transverse--traceless tensorial modes associated to the gravitational sector and the transverse vectorial modes associated to the gauge vector interaction satisfy independent linear wave equations, without any coupling between them. The Newtonian part of the anisotropic theory, very relevant near the sources, does not affect the free propagation of the physical degrees of freedom in the wave zone. It turns out that both excitations, the gravitational and the vectorial one, propagate with the same speed $\sqrt{\beta}$, where $\beta$ is the coupling parameter of the scalar curvature of the three dimensional leaves of the foliation defining the Ho\v{r}ava--Lifshitz geometry.

\end{abstract}
%\date{}
\maketitle

\section{Introduction}
Due to recent detection of gravitational waves (GW), last years the multimessenger astronomy has increased in recent years. This has allowed the exploration of our Universe in a deeper way\cite{Miller2019,Abbottetal2021gwtc3,Nitzetal2021}.

GW's events together with its electromagnetic counterpart detection, generated by neutron stars coalescence, have enabled to compare the gravitational and electromagnetism waves speed propagation \cite{AbbottEtal2017,AbbottEtal2017a}. 

It is to be expected that as detection techniques improve -whether with new experiments, redesigns or improvements in current data analysis techniques- a better understanding of these phenomena will occur and it could be determine if there is any discrepancy between experimental data and successful theoretical models from General Relativity (GR) \cite{BailesetAl2021}.   

On the other hand, astroparticle experiments by means of ultra high energy (UHE) photons coming from gamma ray burst (GRB), are exploring the Universe on the largest energy scales observed so far. Particularly, the LHAASO experiment has detected on the Earth's surface, gamma rays at the PeV energy scale \cite{cao2021ultrahigh,LiMa2021a,xu2016(1),xu2016(2),xu2018,Martinez-Huerta-2020}. These detections are not in agreement with those models where the photon relation dispersion is of the form 
$\omega^{2}\propto k^{2}$. These experiments suggest that at a certain energy scale of order of $3,6\times 10^{17}$ GeV, speed of propagation of the photons depends on the energy scale, hence one needs a modified dispersion relation (MDR) which implies a Lorentz invariance violation (LIV) \cite{LiMa2021a}. It has been argue that this observations could be consistent with String/ M--theory inspired quantum gravity \cite{LiMa2021a,LiMa2021b},
In this work we will argue in terms of a model of anisotropic gravity, proposed by Ho\v{r}ava, coupled to a gauge vector, describing the electromagnetic interaction.

Since the Ho\v{r}ava proposal \cite{Horava2009} and its consistent extension \cite{BlasPojolasSibiryakov2010}, many investigations have been devoted to corroborated the consistency of the theory as a possible candidate to describe a UV complete gravitational theory \cite{BlassPujolassibiryakov2010,CharmousisEtal2009,PapazoglouSotiriousThomas2010,OrlandoReffert2009,Kluson:2010nf,Bellorin:2010te,Donnelly:2011df,Bellorin:2011ff,Bellorin:2012di,ContilloRechenbergerSaueressig2013,BenedettiGuarnieri2014,DOdoricoSaueressingSchutten2014,DOdoricoEtal2015,BarvinskyEtal2016}. Furthermore, the connection with other relevant theories has been established. For example, the Ho\v{r}ava--Lifshitz theory at low energy coincides with the Einstein--aether theory \cite{Jacobson:2000xp} under the hypersurface orthogonal gauge condition \cite{Jacobson:2010mx,Jacobson:2013xta}. Also, it has been related with Causal Dynamical Triangulations \cite{Ambjorn:2010hu} and the Newton--Cartan theory \cite{Hartong:2015zia}, to name a few.

In a broader context, the behavior of this anisotropic theory of gravity when coupled to matter fields has been studied in \cite{Kimpton:2013zb,PospelovShang2012,Colombo:2014lta,Colombo:2015yha}. An alternative proposal was considered in \cite{BellorinRestucciaTello2018b,RestucciaTello2020a}, where the anisotropic gravity--gauge vector coupling in 3+1 dimensions arises from a higher dimensional Ho\v{r}ava--Lifshitz theory at the kinetic conformal point through a dimensional reduction scheme. In this way, a consistent formulation can be obtained where only the gravitational and vectorial degrees of freedom propagate, no scalar degree of freedom is present. The corresponding field equations, in the low energy regime, depend on two couplings $\alpha$ and $\beta$. According to the known experimental data $\alpha$ must be very near 0 and $\beta$ very near 1. It turns out that the field equations of this formulation evaluated at $\alpha=0$ and $\beta=1$ agree exactly with the Einstein--Maxwell equations in a particular gauge (the gravitational gauge \cite{ArnowitDesrMisnert2008}). This is an important limit for the Ho\v{r}ava--Lifshitz at the kinetic conformal point (KC) proposal \cite{BellorinRestucciSotomayor2013,BellorinRestuccia2016B}. Moreover, both theories, the described anisotropic Ho\v{r}ava--Lifshitz and the Einstein--Maxwell one, propagate for any value of the coupling parameters the same degrees of freedom, the transverse traceless tensorial modes (the TT--modes) for the gravitational sector and the transverse modes for the gauge vector (the T--modes).

Ho\v{r}ava's proposal also has been challenged in the cosmological (within the framework of the projectable version) and experimental scenarios. For the former, some problems such as the phase space for a wide range of self--interacting potentials for the scalar field \cite{Leon:2019mbo}, bouncing cosmology for entropy corrected models \cite{Bandyopadhyay:2019xbv} and the topological classification of the Universe space--time when the cosmological constant is absents, have been explored. Static and spherically symmetric solutions of the Ho\v{r}ava--Lifshitz field equations in the context of the non--projectable version have been found. Particularly, these solutions correspond to wormhole throats \cite{Bellorin:2014qca,Bellorin:2015oja,RestucciaTello2021}, although in the projectable version \cite{Garcia-Compean:2020aaa} and in 2D \cite{Ambjorn:2021wou} wormholes structures were obtained.

Concerning the experimental challenges, recent detection of gravitational waves \cite{AbbottEtal2017,AbbottEtal2017a} restricts the coupling parameter $\alpha$ and $\beta$ (both appearing in the potential in the low energy regime $z=1$). Using the data of the GW170817 they find $\{\beta\sim 1, \alpha\sim 0\}$ \cite{EmirGumrukcuogluSaravaniSotiriou2018,RamosBarausse2019,Barausse:2019yuk}, what is more using the data of GRB170817A in a FLRW background in \cite{Zhang:2020bzg} the authors determined that $|1-\sqrt{\beta}|<\left(10^{-19}-10^{-18}\right)$, thus $\beta\sim 1$.

In this regard, the generation of gravitational waves in the Ho\v{r}ava--Lifshitz theory and its properties were analyzed in \cite{Blas:2011zd}. Recently, the existence of a wave zone at both the IR energy regime \cite{MestraPenaRestuccia2021} and the UV energy regime \cite{MestraPenaRestuccia2021a} was proven. 
In this zone, far away from the sources, the physical degrees of freedom, the transverse--traceless modes (TT) at leading order $\mathcal(1/r)$, propagate following a linear wave equation. Although the equations for this modes coincide with the field equations they satisfy at linearized level around a Minkowski background with propagation speed $\sqrt{\beta}$, there are also nontrivial Newtonian metric components not present in the linearized formulation. Some of this Newtonian terms are of the same order $\mathcal(1/r)$ but do not contribute to the wave equations. In particular, they contribute to the gravitational energy and are relevant in the near zone to the sources and to the asymptotic behavior 
of the gravitational field \cite{MestraPenaRestuccia2021,MestraPenaRestuccia2021a,ArnowittDesserMisner1961}.

Taking into account these interesting antecedents, the main aim of the present investigation, is to determine the wave zone in the pure anisotropic gravity--gauge vector model in the framework of the non--projectable Ho\v{r}ava--Lifshitz theory at the KC point, at low energy scale. A natural goal is to find the MDR for this model  and compare it with the astrophysical experiments for UHE photons. This goal requires a complete understanding of the wave zone, at all energy scales. In the present article we will prove that, at low energies and taking into account all nonlinearities of the theory, both the gravitational and gauge vector waves propagate freely, without any coupling between them,
in certain region of the space, the wave zone. We will compare it with the wave zone in the Einstein--Maxwell theory. The analysis at high energy will be discussed elsewhere.  

The starting point shall be the model given in \cite{RestucciaTello2020a} and we will follow the approach in \cite{MestraPenaRestuccia2021,MestraPenaRestuccia2021a}. The wave zone will be defined and, after solving all constraints and field equations for both the gravitational and vector sectors subject to a suitable gauge condition, the contribution of the physical degrees of freedom and the source term will be determined. Moreover, we will obtain the full energy of the system where the non--trivial Newtonian components of the metric are quite involved. We will show that both the TT--modes for the gravitational part and the T--modes for the gauge vector sector propagate satisfying wave equations without any interaction with the non--trivial Newtonian background as it occurs in GR. Therefore, the present study constitutes a first step in the analysis of the existence of a wave zone in the case of material coupling to Ho\v{r}ava--Lifshitz gravity, which is non--trivial given the presence of crossed terms between the two sectors, which can destroy the conditions required for a well--established wave zone.

The article is organized as follows: Sec. \ref{sec2} presents the pure anisotropic gravity--gauge field coupling, its fields equations and constraints. Besides, the full energy of the system is determined. In Sec. \ref{sec3} the wave zone hypothesis is presented. Also the field equations and the full set of constraints are solved in order to estimate the behavior of the physical degrees of freedom in the wave zone. Finally, Sec. \ref{sec4} concludes the investigation.

\section{The pure anisotropic gravity--vector gauge coupling theory}\label{sec2}

As we are interested in studying both, the gravitational and electromagnetic waves in the framework of the non--projectable Ho\v{r}ava--Lifshitz theory, the starting point is the Hamiltonian at low energies ($z=1$) of the pure anisotropic gravity--vector gauge coupling given by \cite{BellorinRestucciaTello2018b,RestucciaTello2020a}
\begin{eqnarray}\label{Hamiltonian}
H&=&\int_{\Sigma_{t}} d^{3}x\bigg\{N\sqrt{g}  \left[ 
	\frac{\pi^{ij}\pi_{ij}}{g}+\frac{E^{i}E_{i}}{2g}-\beta R + \frac{\beta}{4}F_{ij}F^{ij}-\alpha a_{i}a^{i} \right]  -\Lambda \tilde{H} -\Lambda_{j}H^{j}-\sigma P_{N}-\mu \pi \bigg\} + \beta \mathcal{E}.
	\end{eqnarray}
It should be noted that the above Hamiltonian is invariant under foliation--preserving diffeomorphisms and a gauge symmetry group \i.e., $\text{FDiff}_{\mathcal{F}}\times \text{U}(1)$ \cite{BellorinRestucciaTello2018b,RestucciaTello2020a}.
The coupling between the anisotropic Ho\v{r}ava--Lifshitz gravity and the $U(1)$ gauge vector is the natural one. The coupling parameter $\beta$ is the same, since (\ref{Hamiltonian}) arises from a dimensional reduction of a higher dimensional Ho\v{r}ava--Lifshitz gravity model. We notice that we can add to the Hamiltonian a term $ \frac{\pi^{2}}{2}$ without modifying the dynamics of the theory, due to the presence of the $\mu\pi$ term, $\mu$ being a Lagrange multiplier. In this form, aside the $\alpha$ and $\mu$ dependent terms the Hamiltonian (for $\beta=1$) is the same as the Einstein--Maxwell Hamiltonian, where $N$ is considered as a canonical variable, with canonical conjugate momentum $P_{N}$.

As usual, $\pi^{ij}\equiv\frac{\partial \mathcal{L}}{\partial \dot{g}_{ij}} $ corresponds to the conjugate momentum of the Riemannian 3--metric $g_{ij}$ whilst $E^{i}\equiv \frac{\partial \mathcal{L}}{\partial \dot{A}_{i}}$ is the momentum of the gauge vector field $A_{i}$. Besides, $R$ is the 3--dimensional or spatial Ricci scalar, $a_{i}=\partial_{i}\text{Ln}N$ the acceleration and $F_{ij}\equiv \partial_{i}A_{j}-\partial_{j}A_{i}$ is the field strength of the  gauge vector field. These terms determine the potential of the theory in the IR regime ($z=1$). The functions $\{\Lambda, \Lambda_{i}, \sigma, \mu\}$ are Lagrange multipliers associated to the primary constraints
\begin{eqnarray}
\label{H-tilde_constraint}
\tilde{H}&\equiv&\partial_{i}E^{i}=0, \\ 
\label{Hj_constraint}
H^{j}&\equiv& 2\nabla_{i}\pi^{ij}+E^{i}g^{jk}F_{ik}=0,\\
\label{pi_constraint}
\pi&\equiv& g_{ij}\pi^{ij}=0, \\ 
\label{PN_constraint}
P_{N}&=&0.
\end{eqnarray}
The boundary term 
\begin{equation}\label{EEE}
  \mathcal{E}=-\beta\oint_{\partial\Sigma_{c}} dS_{i}\left(\partial_{j}g_{ij}-\partial_{i}g_{jj}\right),
\end{equation}
ensures the Frechet differentiability of the action. It was introduced in GR by \cite{ReggeTeitelboim1974} and proven to be the ADM gravitational energy.
Note that (\ref{H-tilde_constraint}) is the same first class constraint associated to the $U(1)$ gauge symmetry group as in the relativistic electromagnetic theory, this corresponds to the Gauss's law. The constraint (\ref{Hj_constraint}) is the generator of space--like diffeomorphisms on the spatial leaf of the foliation. $\tilde{H}$ and $H^{j}$ are first class constraints, being (\ref{pi_constraint}) and (\ref{PN_constraint}) second class constraints. The conservation of these second class constraints provides 
\begin{equation}
\label{HP_constraint}
\begin{split}
H_{P}\equiv\frac{3}{2}\frac{1}{\sqrt{g}}\pi^{lm}\pi_{lm}+\frac{1}{4}\frac{1}{\sqrt{g}}E^{k}E_{k}+\frac{1}{2}\sqrt{g}\beta R+\frac{1}{8}\sqrt{g}\beta F^{lm}F_{lm}+\sqrt{g}\left(\frac{\alpha}{2}-2\beta\right)a^{k}a_{k}  -2\beta\sqrt{g}\nabla^{l}a_{l}=0,
\end{split}
\end{equation}

\begin{equation}
\label{HN_constraint}
\begin{split}
H_{N}\equiv\frac{1}{\sqrt{g}}\bigg[\pi^{ij}\pi_{ij}+\frac{E^{i}E_{i}}{2}-\beta g R +\frac{\beta}{4} g F_{ij}F^{ij} \bigg]+\alpha\sqrt{g}a_{i}a^{i}
+2\alpha\sqrt{g}\nabla_{i}a^{i}=0.
\end{split}
\end{equation}
The expressions (\ref{HP_constraint}) and (\ref{HN_constraint}) also are of second class,
up to a suitable integral combination of the constraints  (\ref{pi_constraint})--(\ref{HN_constraint}), a first class constraint generator of time reparametrizations \cite{tesis}.

Now, the field equations describing the dynamics of the anisotropic gravity and gauge vector sectors, are obtained by taking variations of (\ref{Hamiltonian}) with respect $\{g_{ij}, \pi^{ij}\}$ and $\{A_{i}, E^{i}\}$ respectively, leading to
\begin{eqnarray}\label{gij_punto_EoM}
\dot{g}_{ij}&=&\frac{2N}{\sqrt{g}}\pi_{ij}+\nabla_{i}\Lambda_{j}+\nabla_{j}\Lambda_{i}+\mu g_{ij}, \\ \label{Ai_punto_EoM}
\dot{A}_{i}&=&\frac{N}{\sqrt{g}}E_{i}+\partial_{i}\Lambda-\Lambda_{j}g^{jk}F_{ik},
\end{eqnarray}
\begin{equation}\label{pi_ij_punto_EoM}
\begin{split}
\dot{\pi}^{ij}=\frac{N}{2}\frac{g^{ij}}{\sqrt{g}}\left[\pi^{lk}\pi_{lk}+\frac{E^{l}E_{l}}{2}\right]  
-\frac{N}{\sqrt{g}}\left[2\pi^{ik}\pi^{j}_{k}+\frac{E^{i}E^{j}}{2}\right]
+N\sqrt{g}\beta\left[\frac{R}{2}g^{ij}
- R^{ij}\right]+\beta\sqrt{g}\bigg[\nabla^{(i}\nabla^{j)}N-g^{ij}\nabla_{k}\nabla^{k}N\bigg]
&\\+\frac{\beta}{2}N\sqrt{g}\Bigg[F^{in}F^{\ j}_{n}
-\frac{g^{ij}}{4}F_{mn}F^{mn}\Bigg]
+\alpha N\sqrt{g} \bigg[\frac{g^{ij}}{2}a_{k}a^{k}
-a^{i}a^{j}\bigg]
-\nabla_{k}\bigg[2\pi^{k(i}\Lambda^{j)}-\pi^{ij}\Lambda^{k}\bigg]-\Lambda^{(i}g^{j)m}E^{l}F_{lm}+\mu \pi^{ij},
\end{split}    
\end{equation}

\begin{equation}\label{Ei_punto_EoM}
\dot{E}^{i}=\beta\partial_{j}\left(N\sqrt{g} F^{ji}\right)+\partial_{k}\left(\Lambda^{k}E^{i}-\Lambda^{i}E^{k}\right).
\end{equation}
As can be appreciated, the above fields equations involve two coupling constants, namely $\beta$ and $\alpha$ and a Lagrange multiplier $\mu$, compared to Einstein--Maxwell theory. The coupling $\beta$ is related with the propagation speed of the tensorial and vector modes associated with the anisotropic gravity and gauge vector. Besides, the expressions describing the dynamic of the gauge vector, that is, (\ref{H-tilde_constraint}), (\ref{Ai_punto_EoM}) and (\ref{Ei_punto_EoM}) only involve the coupling $\beta$. It turns out that, if $\beta=1$ and $\alpha =0$ all the field 
equations are exactly the same ones as in the Einstein--Maxwell relativistic theory, in a particular gauge.  In fact, under the assumption $\alpha=0$ and $\beta=1$, the conservation of the constraint 
$H_{N}=0$ using (\ref{HN_constraint}), (\ref{gij_punto_EoM}), (\ref{Ai_punto_EoM}) and (\ref{pi_ij_punto_EoM}) determines $\mu=0$. The field equations then reduce to the Einstein-Maxwell field equations in the gauge $\pi=0$. The reduced constraint (\ref{HP_constraint}) arises from the Einstein--Maxwell evolution equations through the conservation of the gauge condition.

\subsection{The energy of the pure anisotropic gravity--gauge field interaction}

As can be seen, the Hamiltonian (\ref{Hamiltonian}) is not a sum of constraints as it is in GR \cite{ArnowitDesrMisnert2008}. Nevertheless, it can be rewritten in terms of the function $H_{N}$ plus an additional term which turns out to be a total divergency. Moreover, we will show in this section that on the constrained submanifold one can perform a canonical reduction to a formulation solely in terms of the physical degrees of freedom, that is the transverse--traceless tensorial modes plus the transverse vectorial ones. Furtheremore, we will evaluate it on a particular gauge and obtain the physical energy of the anisotropic gravity--gauge vector model on this coordinate system.  \\

In the evaluation of the Hamiltonian, it is necessary to assume the following asymptotic flat behavior \cite{MestraPenaRestuccia2021} 
\begin{equation}
\label{asymptotic_conditions}
\begin{aligned}
g_{i j}-\delta_{i j}&=\mathscr{O}(1 / r), & \partial g_{i j} &=\mathscr{O}\left(1 / r^{2}\right), \\
\pi^{i j}&=\mathscr{O}\left(1 / r^{2}\right), & \partial \pi^{i j} &=\mathscr{O}\left(1 / r^{3}\right), \\
N-1 &=\mathscr{O}(1 / r), & \partial N &=\mathscr{O}\left(1 / r^{2}\right), \\
N_{i} &=\mathscr{O}(1 / r), & \partial N_{i} &=\mathscr{O}\left(1 / r^{2}\right),
\end{aligned}
\end{equation}
for the gravitational sector and
\begin{equation}
    A_{i}=\mathscr{O}(1 / r), \quad \partial A_{i}=\mathscr{O}(1 / r^{2}),
\end{equation}
for the gauge vector. Additionally, it is convenient to use the following T+L decomposition \cite{MestraPenaRestuccia2021}
\begin{equation}\label{C}
    f_{ij}=f^{T\tau}_{ij}+f^{\tau}_{ij}+\frac{1}{3}\delta_{ij}f,
\end{equation}
where 

\begin{equation}
  \delta_{ij}f^{T\tau}=\delta_{ij}f^{\tau}_{ij}=0, \quad \partial_{i}f^{T\tau}_{ij}=0 
\end{equation}
and
\begin{equation}
    f^{\tau}_{ij}\equiv \partial_{i}W_{j}+\partial_{j}W_{i}-\frac{2}{3}\delta_{ij}\partial_{k}W_{k}.
\end{equation}
This decomposition, assuming appropriate boundary conditions, exists and it is unique. It is analogous to the York decomposition \cite{York:1973ia}, in contrast it is not covariant. The relation with the ADM T+L decomposition \cite{ArnowitDesrMisnert2008} 
\begin{equation}
    f_{ij}=f^{TT}_{ij}+f^{T}_{ij}+\partial_{i}f_{j}+\partial_{j}f_{i},
\end{equation}
where
\begin{equation}
    f^{T}_{ij}\equiv \frac{1}{2}\left(\delta_{ij}f^{T}-\frac{\partial_{i}\partial_{j}}{\Delta}f^{T}\right), \quad \Delta\equiv\partial_{i}\partial_{i},
\end{equation}
is the following
\begin{equation}
    f^{T\tau}_{ij}=f^{TT}_{ij}, \quad W_{i}=f_{i}-\frac{1}{4\Delta}\partial_{i}f^{T}, \quad f=f_{ii}=f^{T}+2\partial_{i}f_{i}.
\end{equation}
The decomposition (\ref{C}) is convenient when using the gauge fixing condition for $g_{ij}=\delta_{ij}+h_{ij}$
\begin{equation}\label{D}
    h_{i}-\frac{1}{4\Delta}\partial_{i}h^{T}=0,
\end{equation}
also used in GR \cite{ArnowitDesrMisnert2008}. In this case
we have
\begin{equation}
    h_{ij}=h^{TT}_{ij}+\frac{1}{2}\delta_{ij}h^{T}=h^{T\tau}_{ij}+\frac{1}{3}\delta_{ij}h.
\end{equation}
This gauge fixing condition is associated to the diffeomorphisms on the spacelike leaves of the Ho\v{r}ava--Lifshitz foliation. \\

The Hamiltonian (\ref{Hamiltonian}) evaluated on the submanifold of the constraints $\Gamma_{c}$ is given by 
\begin{equation}
    H\bigg{|}_{\Gamma_{c}}=-2\alpha \int_{\Gamma_{c}}\sqrt{g}\nabla_{i}\nabla^{i}N+\mathcal{E}=-2\alpha\oint_{\partial\Gamma_{c}}dS_{i}\partial_{i}N-\beta\oint_{\partial\Gamma_{c}}dS_{i}\partial_{i}g^{T}.
\end{equation}
We may follow now the same approach as in \cite{MestraPenaRestuccia2021,MestraPenaRestuccia2021a} with the result given there. However, we would like to obtain explicit expressions in terms of the gravitational and gauge vector contributions. We shall then give the explicit expression of the energy when we consider up to quadratic expressions on the fields. Starting with the constraint $H_{N}=0$ and $H_{p}=0$. To first order on the fields we obtain, using the gauge (\ref{D}),
\begin{eqnarray}
\beta \Delta g^{T}+2\alpha \Delta N&=&0, \\
-\frac{1}{2}\beta \Delta g^{T}-2\beta \Delta N&=&0,
\end{eqnarray}
hence $g^{T}=0$, $N=1$ to first order provided $\beta\neq 0$ and $\alpha-2\beta \neq 0$ (experimental data restricts $\beta$ to be very near to 1 and $\alpha$ to 0 \cite{RamosBarausse2019}). We now evaluate $\beta\Delta g^{T}+2\alpha \Delta N$ to second order on the fields. We get from $H_{N}=0$ and using 
\begin{equation}
    R\cong -\Delta g^{T}-\frac{1}{4}\partial_{k}g^{TT}_{ij}\partial_{k}g^{TT}_{ij}+\text{total divergence},
\end{equation}
where the total divergence, under the assumed boundary conditions, when integrated on the spacelike leaves of the foliation vanishes. We then obtain from the constraint $H_{N}=0$, up to second order on the fields, 
\begin{equation}\label{F}
    -\beta \Delta g^{T}-2\alpha \Delta N=\mathcal{H}+\text{total divergence,}
\end{equation}
where where the Hamiltonian density $\mathcal{H}$ is given by 
\begin{equation}
    \mathcal{H}=\pi^{TTij}\pi^{TTij}+\frac{1}{2}E^{Ti}E^{Ti}+\frac{1}{4}\beta \partial_{k}g^{TT}_{ij}\partial_{k}g^{TT}_{ij}+\frac{\beta}{4}F^{ij}F_{ij}.
\end{equation}
Besides, we have used that, to first order, $g^{T}$ and $N-1$ vanish. Also we may combine $H_{p}=0$ and $H_{N}=0$ to obtain 
\begin{equation}
  \left(2\beta-\alpha\right)\Delta N=  2\pi^{ij}\pi_{ij}+\frac{1}{2}E^{i}E_{i}+\frac{\beta}{4}F^{ij}F_{ij}.
\end{equation}
We have to impose also the other constraints. We get to first order 
\begin{eqnarray}
\pi^{ii}&=&0,\\
\partial_{i}\pi^{ij}&=&0.
\end{eqnarray}
They imply that 
\begin{equation}
    \pi^{ij}=\pi^{TTij}.
\end{equation}

Also, from the constraint (\ref{H-tilde_constraint})
\begin{equation}
    E^{i}=E^{Ti},
\end{equation}
and $A_{i}$ appears only through 
\begin{equation}
    F_{ij}=\partial_{i}A_{j}-\partial_{j}A_{i}=\partial_{i}A^{T}_{j}-\partial_{j}A^{T}_{i}.
\end{equation}
The independent fields become $g^{TT}_{ij}$, $\pi^{TTij}$, $A^{T}_{i}$ and $E^{Ti}$, all other fields been determined in terms of them. The Lagrangian $L$ becomes then 
\begin{equation}
    L\bigg{|}_{\Gamma_{c}}=\int_{\Gamma_{c}}d^{3}x\left(\pi^{ij}\dot{g}_{ij}+E^{i}\dot{A}_{i}-\mathcal{H}\right),
\end{equation}
but 
\begin{equation}
    g_{ij}=g^{TT}_{ij}+\frac{1}{2}\delta_{ij}g^{T},
\end{equation}
hence
\begin{equation}
\pi^{ij}\dot{g}_{ij}=\pi^{TTij}\dot{g}^{TT}_{ij} \quad \mbox{and} \quad E^{i}\dot{A}_{i}=E^{Ti}\dot{A}^{T}_{i}+\text{total divergence}.    
\end{equation}
Finally we get the Lagrangian $L$ to be
\begin{equation}
\label{L_gammac}
    L\bigg{|}_{\Gamma_{c}}=\int_{\Gamma_{c}}d^{3}x\left( \pi^{TTij}\dot{g}^{TT}_{ij}+E^{Ti}\dot{A}^{T}_{i}-\mathcal{H}\right).
\end{equation}

We can then interpret, at quadratic order on fields, $H_{N}=0$ as the Hamiltonian constraints, since we obtain from it the energy density $\mathcal{H}$. In GR, besides the gauge condition (\ref{D}), one can impose a gauge condition to determine the time coordinate. One can choose \cite{ArnowitDesrMisnert2008} 
\begin{equation}\label{tt}
    t=\frac{1}{2\Delta}\pi^{ii},
\end{equation}
corresponding to the gauge condition
 $\pi^{ii}=0$. In fact, $\Delta t=0$.

In the Ho\v{r}ava--Lifshitz formulation we are not allowed to fix (\ref{tt}), since the Hamiltonian constraint is a second class one. Therefore, the time coordinate must be determined from a different argument. \\

In the anisotropic model, to first order 
\begin{equation}
    \pi^{ii}=0.
\end{equation}
We may then find the solution of 
\begin{equation}
    \Delta \phi=\pi^{ii},
\end{equation}
assuming $\phi$ asymptotically constant as a spacelike function. We then get $\phi=f(t)$, and now use the residual gauge symmetry, that is, the reparametrization on time to obtain 
\begin{equation}
    \phi=-2\beta t,
\end{equation}
hence
\begin{equation}\label{H}
    t=-\frac{1}{2\beta\Delta}\pi^{ii},
\end{equation}
is the time coordinate in the anisotropic formulation.The particular coefficient $-2\beta$ arises in order to have $N=1$ up to order 1. In fact, from Eq. (\ref{pi_ij_punto_EoM}) we get
\begin{equation}
    -\frac{1}{2\beta\Delta}\dot{\pi}^{ii}=N+\text{second order terms.}
\end{equation}
Consequently, up to first order 
\begin{equation}
    -\frac{1}{2\beta\Delta}\dot{\pi}^{ii}=1,
\end{equation}
in agreement with (\ref{H}).

Finally,  we may compare with the result of the general and exact expresion for the kinematic term obtained in \cite{MestraPenaRestuccia2021}. From eq.(35)  in \cite{MestraPenaRestuccia2021} we obtain the followings expressions, under the gauge condition (\ref{D}),

\begin{equation}
    \pi^{ij}\partial_{t}h_{ij}= \left(1+ \frac{1}{2}h^{T}\right)\pi^{ij} \partial_{t}\left(\frac{h_{ij}^{T\tau}}{1+\frac{1}{2}h^{T}}\right),
\end{equation}
 since $\frac{1}{3}h=\frac{1}{2}h^{T}$. $h_{ij}^{T\tau}$ and $h$ were defined in (\ref{C}) and $\pi_{ij}$ expresed in terms of the covariant York \cite{York:1973ia} decomposition becomes 
 \begin{equation}
     \pi^{ij}=\tilde{\pi}^{ij T\tau} +\tilde{\pi}^{ij \tau},
 \end{equation}
 $\tilde{\pi}^{ij \tau}$ satisfies
 
 \begin{equation}
 \label{B}
     \nabla_{i}\tilde{\pi}^{ijT\tau}=0, \,\,\, g_{ij}\tilde{\pi}^{ijT\tau}=0,
 \end{equation}
while 

\begin{equation}
    \tilde{\pi}^{ij\tau}= \nabla^{i}U^{j}+\nabla^{j}U^{i}-\frac{2}{3}g^{ij}\nabla_{k}U^{k}.
\end{equation}

The constraints (\ref{Hj_constraint}),(\ref{HP_constraint}) and (\ref{HN_constraint}) determine the well posed elliptic equations wich determine, under the asymptotic boundary conditions (\ref{asymptotic_conditions}), unique solution for $U^{i}$, $h^{T}$ and $N$ as functionals of $h_{ij}^{TT}$, $\pi^{ijTT}$, $E^{iT}$ and $A_{j}^{T}$. The constraint (\ref{H-tilde_constraint})  reduce $E^{i}$ to its transverse part $E^{iT}$. Besides, eq.(\ref{B}) using the argument in \cite{MestraPenaRestuccia2021} eq(34) determines $\tilde{\pi}^{ijT\tau}$ as a functional of the same degrees of fredom  $h_{ij}^{TT}, \,\, \pi^{ijTT}, \,\, E^{T}, \,\, \text{and}\,\,A_{j}^{\tau}$, consequently,  the kinetic term on the constrained submanifold and under the coordinate condition (\ref{D}) is given by 
\begin{equation}
    \pi^{ij}\partial_{t}h_{ij}+E^{i}\partial_{t}A_{i}=\left(1+\frac{1}{2}h^{T}\right)\left(\tilde{\pi}^{ijT\tau}+\tilde{\pi}^{ji\tau}\right)\partial_{t}\left(\frac{h_{ij}^{T\tau}}{1+\frac{1}{2}h^{T}}\right)+E^{iT}\partial_{t}A_{i}^{T}. 
\end{equation}

When we consider only the quadratic expression on the fields, it follows that up  to first order 

\begin{equation}
 \left(1+\frac{1}{2}h^{T}\right)\left(\tilde{\pi}^{ijT\tau}+\tilde{\pi}^{ji\tau}\right)\cong \pi^{ijTT},     
\end{equation}
and 
\begin{equation}
\frac{h_{ij}^{T\tau}}{1+\frac{1}{2}h^{T}} \cong h_{ij}^{TT},    
\end{equation}
 hence we get, up to second order,  the expression (\ref{L_gammac}).

\section{The pure anisotropic gravity--gauge field wave zone}\label{sec3}

In this section a short review about the wave zone definition in the context of General Relativity is given \cite{ArnowittDesserMisner1961, ArnowitDesrMisnert2008}. This argument is extended within the framework of theories with Lorentz symmetry breaking, specifically in the arena of Ho\v{r}ava--Lifshitz gravity theory (for further details see \cite{MestraPenaRestuccia2021,MestraPenaRestuccia2021a}). After that, the wave zone hypothesis for the gauge vector sector are imposed. Then, by using the T+L ADM decomposition the constraints are solved in order to estimate the static and oscillatory parts of the variables and the source terms. 

\subsection{The wave zone definition revisited}

The GR wave zone is defined by assuming the following hypothesis \cite{ArnowittDesserMisner1961,ArnowitDesrMisnert2008} 
\begin{itemize}

\item For $kr\gg1$, where $k$ is the wave number and $r$ the radial distance, gradients and derivatives acting on the canonical variables, decay at least as $\mathcal{O}(\tilde{A}/r)$.

\item The components of the metric tensor $g_{\mu\nu}$, deviate from a flat background (Lorentzian metric $\eta_{\mu\nu}$) by small terms compared to unity. Decreasing at least as $\mathcal{O}(1/r)$ in the wave zone. Then $|g_{\mu\nu}-\eta_{\mu\nu}|\ll1$.

\item Finally, $|\partial g_{\mu\nu}/\partial (kr)|^{2}\ll|g_{\mu\nu}-\eta_{\mu\nu}|$ is required for waves with frequencies $k$ to behave as free radiation. 
\end{itemize}

In the wave zone, gravitational canonical TT--modes behave to leading order like 
    \begin{equation}\label{e46}
        \sim \tilde{A}\frac{e^{i\left(\vec{k}\dot\vec{r}-\omega t\right)}}{r}
    \end{equation}
   being $\tilde{A}=\tilde{A}(t,\theta,\phi)$ a generic function of time and angular coordinates, such that $\tilde{A}$ and all its derivatives are
bounded. Besides, the gravitational background is static \i.e., time independent. Being its asymptotically behavior  
\begin{equation}\label{e47}
    \sim \frac{\tilde{B}}{r},
\end{equation}
with $\tilde{B}=\tilde{B}(\theta,\phi)$. It should be noted that, despite the background is the same order of the TT--modes (contrary to what happens the linearized analysis, where the background is vanishing), the latter ones are satisfying a linear wave equation. 

It is worth mentioning that, beyond the wave front, the gravitational canonical modes decay rapidly. This behavior is necessary to guarantee a finite energy.

The first statement, is the same one imposed in linear theories such as Maxwell electromagnetism theory. Notwithstanding, as one is dealing with a non--linear theory, the remaining requirements are strictly necessaries in order to ensure that the self--interaction terms do not destroy the free propagation of the dynamical modes.

Interestingly, it was shown in \cite{MestraPenaRestuccia2021,MestraPenaRestuccia2021a} that the same GR wave zone hypothesis can by applied in the pure anisotropic gravity in the Ho\v{r}ava--Lifshitz framework. Specifically, in the ADM language, the third and fourth statements read as \cite{MestraPenaRestuccia2021}: $|g_{ij}-\delta_{ij}|\sim|N-1|\sim|\Lambda_{i}|=\mathcal{O}(\tilde{A}/r)\ll1$ and $|\partial g_{ij}/\partial(kr)|^{2}\sim|\partial N/\partial(kr)|^{2}$
$\sim|\partial \Lambda_{i}/\partial(kr)|^{2}\ll|g_{ij}-\delta_{ij}|$, respectively.

As we are interested in incorporating a gauge vector field, additionally to previous hypothesis we impose for the gauge sector: $kr\gg 1$, $|A_{i}| =\mathcal{O} (\tilde{A}/r) \ll1$ and $|\partial A_{i}/\partial(kr)|^{2}\ll|A_{i}|$. The statement $kr\gg 1$ is the usual condition given in the electromagnetism Maxwell theory which is valid here. These requirements are necessary, since we expect that the pure anisotropic gravity--gauge vector wave zone exits. Then, both the gravitational TT--modes and the T--modes associated with the gauge sector behave as (\ref{e46}) on a gravitational background like (\ref{e47}).

Therefore, from the above statements one directly obtains the following estimates on the gravitational and gauge sectors 

\begin{eqnarray}
\label{asymptotic-metric-deviation}
g_{i j}-\delta_{i j}\sim \Lambda_{i}\sim N-1\sim A_{i} &\lesssim& \frac{\tilde{B}}{r}+\frac{\tilde{A} e^{ikr}}{r},
	\end{eqnarray} 
	
	\begin{eqnarray}
	\label{asymptotic-spacial-temporal-metric-spatial-derivative}
	\partial\Lambda_{i}\sim \partial g_{ij} \sim\Gamma^{i}_{\hspace{0.1cm}jl}\sim F_{ij} \lesssim \frac{\tilde{B}}{r^{2}}+k\frac{\tilde{A} e^{ikr}}{r}\,.
	\end{eqnarray}\\
In obtaining (\ref{asymptotic-spacial-temporal-metric-spatial-derivative}) the hypothesis $kr\gg 1$ has been employed.

	\subsection{Constraints solution}
 Now we are in position to determine the behavior of the propagating parts of the momenta $\pi^{ij}$ in the wave zone. To do this, the primary constraints (\ref{Hj_constraint})--(\ref{HP_constraint}) must be solved using the T+L decomposition \cite{MestraPenaRestuccia2021}. We first estimate the behavior of the transverse $\pi^{Tij}$ and longitudinal $\pi^{Lij}$ parts. So, taking the derivative (and expanding the covariant derivative of the tensor density $\pi^{ij}$) in the constraint (\ref{Hj_constraint}), one gets 
	\begin{equation}
	\label{Lapalcian_pi-i,i}
	\Delta \partial_{j}\pi^{j}=-\frac{1}{2}\partial_{j}\left(\pi^{lm}\Gamma^{j}_{\hspace{0.1cm}lm}+E^{k}g^{ij}F_{ik}\right).
	\end{equation}
The right hand side member of the expression (\ref{Lapalcian_pi-i,i}) shall be denoted as	
\begin{equation}
\label{MomentumDensity}
    \mathcal{P}^{j}\equiv \pi^{lm}\Gamma^{j}_{\hspace{0.1cm}lm}+E^{k}g^{ij}F_{jk} \lesssim \frac{\tilde{B}}{r^{3}}+k^{2}\frac{\tilde{A} e^{ikr}}{r^{2}}\,,
\end{equation}
and corresponds to the source terms. It should be noted that in the pure anisotropic gravity case \cite{MestraPenaRestuccia2021}, the object $F_{ij}$ is absents. Next, using the inversion formulae for the Laplacian \cite{ArnowitDesrMisnert2008,MestraPenaRestuccia2021a} we can estimate the behavior of the solution of (\ref{Lapalcian_pi-i,i}) and then estimate the longitudinal part. Then, the longitudinal part $\pi^{Lij}$ behaves in the wave zone as

\begin{equation}
\label{EstimatePiLongitudinal}
    \pi^{Lij}\lesssim \frac{\tilde{B}}{r^{2}}+k\frac{\tilde{A} e^{ikr}}{r^{2}}\,.
\end{equation}

Applying the same procedure to the constraint (\ref{pi_constraint}), this yields $\pi^{T}+2\pi^{i}_{\hspace{0.1cm},i}=0$. Hence, the estimation of the transverse part $\pi^{Tij}$ of the momentum behavior in the wave zone reads as 
\begin{equation}
\label{EstimatePiTrasnverso}
    \pi^{Tij}\lesssim \frac{\tilde{B}}{r^{2}}+k\frac{\tilde{A} e^{ikr}}{r^{2}}\,. 
\end{equation}
As can be appreciated, both the longitudinal and transverse parts of the momentum $\pi^{ij}$ behave as $\mathcal{O}\left(1/r^{2}\right)$. Therefore, only the TT--components of the momentum might behave as $\mathcal{O}\left(1/r\right)$. The solution of the remaining constraints and the analysis of the dynamical equations allow to find out the appropriate behavior of $\pi^{ijTT}$ in the wave zone.
In fact, we obtain
\begin{equation}
\pi^{TTij} \lesssim \frac{B^{i j}}{r}+k \frac{\tilde{A}^{i j} e^{i k r}}{r}.
\end{equation}
On the other hand, it can be shown from the constraint (\ref{H-tilde_constraint}) and the T+L decomposition, that the longitudinal part of the momentum $E^{Li}$ of the gauge sector is zero \cite{BellorinRestucciaTello2018b,RestucciaTello2020a}.\\

Now, to infer the behavior of the remaining canonical variables \i.e., $g_{ij}$ and $N$, we need to use the second class constraints (\ref{HP_constraint}) and (\ref{HN_constraint}) together with the previous estimation on the components of the momentum $\pi^{ij}$. The suitable combination $2H_{p}+H_{N}$ of the constraints (\ref{HP_constraint}) and (\ref{HN_constraint}) leads to 
\begin{equation}
    \Delta N \lesssim \frac{\tilde{B}}{r^{2}}+k^{2}\frac{\tilde{A} e^{ikr}}{r^{2}}\,,
\end{equation}
where the condition $\alpha\neq 2\beta$ should be taken into account. It is remarkably to note that these objects are zero when the linearized treatment is performed \cite{BellorinRestucciSotomayor2013,BellorinRestuccia2016B}, as occurs in the GR case \cite{ArnowitDesrMisnert2008,ArnowittDesserMisner1961}. However, as pointed out before, in the asymptotic scheme these objects have a non--trivial Newtonian--like part $\mathcal{O}\left(1/r\right)$.
Then, using the transverse gauge condition $g_{ij,j}=0$ on constraints (\ref{HP_constraint}) and (\ref{HN_constraint}) one arrives to
  \begin{equation}
    \label{N-estimate}
    N -1\sim g^{T}\sim\frac{B}{r} +\frac{\tilde{A}  e^{ikr}}{r^{2}} \,.  
\end{equation}
 With the above information, the behavior of the vector $a_{i}$ and consequently its contribution in the low energy regime to potential of the theory (the object $\alpha a_{i}a^{i}$ breaks manifestly
the relativistic symmetry) is estimated, given the results 
  \begin{eqnarray}
	a_{i}\lesssim \frac{B_{i}}{r^2}+  k \frac{\tilde{A}_{i}  e^{ikr}}{r^{2}}, \\
	\label{a^{1}a_{i}}
	a_{i} a^{i}\lesssim \frac{B}{r^{4}}+ k^{2} \frac{\tilde{A}  e^{ikr}}{r^{4}},\\
	\label{nabla^i a_{i}}
	\nabla^{i}a_{i} \lesssim \frac{B}{r^{3}}+k^{2} \frac{\tilde{A}  e^{ikr}}{r^{2}}\,.
	\end{eqnarray} 
	
From the field equation (\ref{gij_punto_EoM}) one can get valuable information. So, applying the T+L decomposition along with the transverse gauge $g_{ij,j}=0$, the equation (\ref{gij_punto_EoM}) becomes 
\begin{eqnarray}
	\label{gpunto-TT}
		\dot{g}_{ij}^{TT}-2 \pi_{ij}^{TT}+ \left(\mu g_{ij}\right)^{TT} &\lesssim&\frac{B}{r^{2}}+k^{2}\frac{\tilde{A}  e^{ikr}}{r^{2}},\\
		\label{gpunto-L}
		-2 \partial_{(j}\Lambda_{i)} + \left(\mu g_{ij}\right)^{L} &\lesssim&\frac{B}{r^{2}}+k^{2}\frac{\tilde{A} e^{ikr}}{r^{2}},\\
		\label{gpunto-T}
		\dot{g}_{ij}^{T}+\left(\mu g_{ij}\right)^{T}&\lesssim&\frac{B}{r^{2}}+k^{2}\frac{\tilde{A}  e^{ikr}}{r^{2}},
	\end{eqnarray}
where terms involving the Lagrange multiplier $\mu$ provide
\begin{eqnarray}
\mu g_{ij}&=&\left(\mu g_{ij}\right)^{TT}+\left(\mu g_{ij}\right)^{T}+\left(\mu g_{ij}\right)^{L}, \\
\left(\mu g_{ij}\right)^{T}&=&\mu \delta_{ij}-\frac{\partial_{i}\partial_{j}}{\Delta}\mu+\mathcal{O}\left(1/r^{2}\right), \\
\left(\mu g_{ij}\right)^{L}&=&\frac{\partial_{i}\partial_{j}}{\Delta}\mu+\mathcal{O}\left(1/r^{2}\right), \\
\left(\mu g_{ij}\right)^{TT}&=&\mathcal{O}\left(1/r^{2}\right).
\end{eqnarray}	
Now, putting together equations
(\ref{N-estimate}) and (\ref{gpunto-T}) we arrive to  
\begin{equation}
\mu\lesssim\frac{B}{r^{2}}+k^{2}\frac{\tilde{A}  e^{ikr}}{r^{2}}.
\end{equation}
The above estimation on the behavior of the Lagrange multiplier $\mu$, allows to determine from equation (\ref{gpunto-L}) 
\begin{equation}
\label{gpunto_(i,j)}
\partial_{(j}\Lambda_{i)}\lesssim\frac{B}{r^{2}}+k^{2}\frac{\tilde{A}  e^{ikr}}{r^{2}} \,,
\end{equation}
thus, for the Lagrange multiplier $\Lambda_{i}$ one gets 
\begin{equation}
\Lambda_{i}\lesssim \frac{B}{r}+k \frac{\tilde{A}  e^{ikr}}{r^{2}}.
\end{equation}
Therefore (\ref{gpunto-TT}) reads as
 \begin{equation}
	\label{canonica-gpinto-TT}
	\dot{g}_{ij}^{TT}=2 \pi_{ij}^{TT}+\mathcal{O}(1/r^{2})\,,
	\end{equation}
	where terms of order of $\mathcal{O}(1/r^{2})$ are given by $\frac{B}{r^{2}}+k^{2}\frac{\tilde{A} e^{ikr}}{r^{2}}$.
	Now, turning back to equation (\ref{pi_ij_punto_EoM}), it is worth mentioning that crossed terms appearing in the right hand member of this expression \i.e., those objects involving gravitational and gauge objects, as well as pure gauge terms such as $\Lambda_{j}g^{jk}F_{ik}$, $E^{l}E_{l}$, $\Lambda^{(i}g^{j)m}E^{l}F_{lm}$,  $F_{nm}F^{mn}$, could in principle destroy the free propagation of the TT--modes. However, the leader order of these term is $\mathcal{O}\left(1/r^{2}\right)$, hence they are not contributing to the $1/r$ order of the propagating modes. Then the free propagation of the TT--modes in the wave zone is saved.
	Thereby, taking into account the last statements, (\ref{pi_ij_punto_EoM}) leads to 
	\begin{equation}
	    \dot{\pi}^{ijTT}=\frac{1}{2}\beta\Delta g_{ij}^{TT}+\mathcal{O}(1/r^{2}).
	\end{equation}
	
	 Finally, for the anisotropic gravitational sector in the leading order, the TT--fields in the wave zone fulfill 
	 \begin{equation}
	     \frac{1}{\beta}\ddot{g}^{TT}- \Delta g_{ij} ^{TT}=0.
	 \end{equation}
This result corroborates that to the leading order $\mathcal{O}(1/r)$ the transverse--traceless tensorial modes propagate satisfying a hyperbolic equation, where the speed propagation is $\sqrt{\beta}$. This wave equation is the same one obtained at linearized level \cite{BellorinRestucciSotomayor2013}. 
Proceeding in an analogous way for the gauge sector, from the equation of motion (\ref{Ai_punto_EoM}) after apply the T+L decomposition and the gauge condition $A_{i}^{L}=0$, one obtains
	 \begin{eqnarray}
	 \label{A_punto_T}
	 \dot{A}_{i}^{T}-E_{i}^{T}&\lesssim& \frac{B}{r^{3}}+k^{2}\frac{\tilde{A} e^{ikr}}{r^{2}}\,,\\
	 \label{A_punto_L}
	 \partial_{i} \Lambda &\lesssim& \frac{B}{r^{3}}+k^{2}\frac{\tilde{A} e^{ikr}}{r^{2}}\,. 
	 \end{eqnarray}
	 From (\ref{A_punto_L}) we can estimate the Lagrange multiplier $\Lambda$ obtaining
	 \begin{equation}
	\Lambda \lesssim  \frac{B}{r^{2}}+k\frac{\tilde{A} e^{ikr}}{r^{3}}.
	 \end{equation}
	 Furthermore, the transverse part of the field equation (\ref{Ei_punto_EoM}) behaves as  
	 \begin{equation}
	     \dot{E}^{Ti}-\beta  A^{Ti}\lesssim \frac{B}{r^{3}}+k\frac{\tilde{A} e^{ikr}}{r^{2}}\,.
	 \end{equation}
	 Then the leading order of the vector gauge field propagate freely in the wave zone too, satisfying the following linear wave equation
	  \begin{equation}
	     \frac{1}{\beta}\ddot{A}_{i}^{T}- \Delta A_{i} ^{T}=0.
	 \end{equation}
As the TT--modes for the gravitational sector, here the transverse T--modes of the gauge sector also propagate with speed $\sqrt{\beta}$, as occurs in the linearized theory \cite{BellorinRestucciaTello2018b,RestucciaTello2020a}. In fact, from the theoretical point of view, in the relativistic theory both the gravitational and electromagnetism radiations propagate with the same velocity, the speed of light. In this regard, recent accurate experimental measurements have determined that theoretical results should match the recent detection of coincident gravitational and electromagnetic radiations \cite{AbbottEtal2017,LIGOScientific:2017ync}. In this concern, the propagation speeds should match to within a part in $10^{15}$ \cite{AbbottEtal2017a,Barausse:2019yuk}. Moreover, using the model proposed in \cite{BellorinRestucciaTello2018b,RestucciaTello2020a}, taking into account the radiation coming from the GRB170817A the value of the coupling $\beta$ was estimated \cite{Zhang:2020bzg}. Concretely the authors have determined that $|1-\sqrt{\beta}|<\left(10^{-19}-10^{-18}\right)$, thus $\beta$ is close to 1. This result agrees with previous investigations using the radiation from GW170817 \cite{EmirGumrukcuogluSaravaniSotiriou2018,RamosBarausse2019}. Then as a primary conclusion, one can say that the pure anisotropic gravity--gauge vector coupling in the framework of the Ho\v{r}ava--Lifshitz theory exits and is well--behaved.

Previous results in GR and in the Ho\v{r}ava--Lifshitz gravity theory at the low and high energy regimes, shows that the propagation of the physical degrees of freedom in the wave zone, up to order $1/r$, fulfill the same equations of motion obtained in the linearized theory. On the other hand, from results given in Refs. \cite{RestucciaTello2020b,tesis}, the evolution equation of the physical degrees of freedom at linear order is   
	  $(-\partial_{tt}+\beta\Delta^{2}+\hat{\beta}_{1}\Delta^{4}-\hat{\beta}_3\Delta^{3}+ \hat{\beta}_{5}\Delta^{4})\psi=0$ for both, the TT--gravitational modes and T-gauge modes sectors, for the $z=4$ case. This result follows directly from a gravity-gauge vector model arising from a dimensional reduction of higher dimensional gravity theory. Therefore, following the result in this article one can expect that the non--relativistic photon in wave zone has the same dispersion relation as the non--relativistic graviton found in \cite{MestraPenaRestuccia2021a} -via  wave zone analysis-
	$\omega^{2}/k^{2}=\beta-\beta_{1}k^{2}-\beta_{3}k^{6}$ 
	 for the $z=3$ case and   $\omega^{2}/k^{2}=\beta-\hat{\beta}_{1}k^{2}-\hat{\beta}_{3}k^{6}-\hat{\beta}_5 k^{8}$ for the $z=4$ case \cite{RestucciaTello2020b,tesis}. This would describe LIV photon whose phase velocity $c(k)\equiv \omega/k$ would depend on the wave number $k$, hence the energy of those photons could be tested using gamma rays data of UHE coming form astrophysical sources such as those detected by
	 LHAASO. 

\section{Concluding remarks}\label{sec4}

In this investigation, we have shown that the wave zone in the non--projectable version of the Ho\v{r}ava--Lifshitz theory at the KC point when coupled with a matter gauge field is well established. As a first step, only those terms that make up the potential in the low energy regime ($z=1$) have been considered. The ADM energy of the system is obtained, and its expression to all orders given. In particular the expression up to quadratic order on the fields is explicitly presented. It depends solely on the TT--gravitational modes and the T--vectorial modes, however the presence of the T--modes of the metric and the lapse function, which are not zero in the wave zone are relevant in the analysis. To obtain the energy, a canonical Hamiltonian reduction to physical degrees of freedom was performed. However, unlike what happens in GR, where the reduction to the physical degrees of freedom (the TT--modes) is through a gauge condition, in the Ho\v{r}ava--Lifshitz theory given its non--covariant character, such gauge choice \emph{per se} does not exist.  Although, the spacelike coordinates in the Ho\v{r}ava--Lifshitz model can be chosen as in GR, the fixing of the time coordinates, in contrast with GR, is not possible. In fact, only reparametrization in the time coordinate is allowed. Even so, the full determination of coordinates in the Ho\v{r}ava--Lifshitz model is explicitly obtained.

The main result of this work, is that the pure anisotropic gravity--gauge vector wave zone exists, where at the leading order $\mathcal{O}(1/r)$ the asymptotic behavior of the of transverse traceless tensorial modes (TT--modes) of the gravitational sector and the transverse modes (T--modes) of the vector part, propagate freely in the wave zone where both are satisfying a wave equation travelling at speed $\sqrt{\beta}$. This result agrees with the linearized theory as happens in the GR case. However, the transverse part of the metric and the lapse function, which are disregard in the linearized analysis, play a preponderant role. This is so because they conform a static Newtonian background which is of the same order as the propagating modes in the wave zone, but it does not interact with them. On the other hand, the TT--components of the metric and the T--modes of the vector sector at large enough distances from the source decay rapidly, since their behavior involves higher powers of $1/r$ and the T--modes and lapse function prevail asymptotically.

Finally, is should be pointed out that the situation involving the full potential of the theory, that is, terms with high order spatial derivatives containing the elliptic operators $\Delta^{2}$, $\Delta^{3}$ and $\Delta^{4}$, shall modify the wave equation of the dynamical degrees of freedom. Also, the Newtonian background might be modified by these terms. It is worth mentioning that the linear operator $\Delta^{4}$ appears as a consequence of the coupling between the anisotropic gravity and the gauge vector field. The study including the complete potential interaction will be addressed elsewhere.   
 
%%%%%%%%%%%%%%%%%%%%%%%%%%%%%%%%%%%%%%%%%%%%%%%%%%%%%%%%%%%%%%%%%%%%%%

\section*{ACKNOWLEDGEMENTS}
 J. Mestra-P\'aez acknowledge financial support by  Beca Doctorado Nacional 2019 CONICYT, Chile.   N° BECA: 21191442. J. Mestra-P\'aez acknowledges the Ph.D. program Doctorado en F\'isica menci\'on en F\'isica Matem\'atica de la Universidad de Antofagasta for continuous support and encouragement. F. Tello-Ortiz thanks the financial support by project ANT-2156 at the Universidad de Antofagasta, Chile.

%%%%%%%%%%%%%%%%%%%%%%%%%%%%%%%%%%%%%%%%%%%%%%%%%%%%%%%%%%%%%%%%%%%%%%%
\bibliographystyle{elsarticle-num}
\bibliography{Bib_Horava_Maxwell}

\end{document}